\documentclass[a4paper,prl,showpacs,superscriptaddress,nofootinbib,twocolumn]{revtex4-1} 
\usepackage[british]{babel}  
\usepackage{lmodern}  
\usepackage[scaled=0.86]{berasans}  
\usepackage[scaled=1.03]{inconsolata} 
\usepackage[usenames,dvipsnames]{color} 
\usepackage[colorlinks,citecolor=blue,linkcolor=magenta,urlcolor=blue]{hyperref}  
\usepackage{graphicx} 
\usepackage[babel]{microtype}  
\usepackage{amsmath,amssymb,amsthm,bm,mathtools,amsfonts,mathrsfs,bbm} 
\usepackage{xspace}  
\usepackage{pgfplots}  

\newcommand{\bra}[1]{\left\langle #1 \right|}
\newcommand{\ket}[1]{\left| #1 \right\rangle}
\newcommand{\braket}[2]{\left\langle #1 \middle| #2 \right\rangle}
\newcommand{\ketbra}[2]{\left|#1\middle\rangle\middle\langle#2\right|}

\newcommand{\mean}[1]{\langle#1\rangle}

\newcommand{\ba}{\begin{eqnarray}}
\newcommand{\ea}{\end{eqnarray}}
\newcommand{\ban}{\begin{eqnarray*}}
\newcommand{\ean}{\end{eqnarray*}}

\usepackage{multirow,bigdelim}

\begin{document}

\title{Covariance Bell Inequalities}

\author{Victor Pozsgay}
\affiliation{D\'epartement de Physique Appliqu\'ee, Universit\'e de Gen\`eve, 1211 Gen\`eve, Switzerland}
\author{Flavien Hirsch}
\affiliation{D\'epartement de Physique Appliqu\'ee, Universit\'e de Gen\`eve, 1211 Gen\`eve, Switzerland}
\author{Cyril Branciard}
\affiliation{Univ. Grenoble Alpes, CNRS, Grenoble INP\footnote{Institute of Engineering Univ. Grenoble Alpes}, Institut N\'eel, 38000 Grenoble, France}
\author{Nicolas Brunner}
\affiliation{D\'epartement de Physique Appliqu\'ee, Universit\'e de Gen\`eve, 1211 Gen\`eve, Switzerland}

\date{\today}  

\begin{abstract}

We introduce Bell inequalities based on covariance, one of the most common measures of correlation. Explicit examples are discussed, and violations in quantum theory are demonstrated. A crucial feature of these covariance Bell inequalities is their nonlinearity; this has nontrivial consequences for the derivation of their local bound, which is not reached by deterministic local correlations. For our simplest inequality, we derive analytically tight bounds for both local and quantum correlations. 
An interesting application of covariance Bell inequalities is that they can act as ``shared randomness witnesses'': specifically, the value of the Bell expression gives device-independent lower bounds on both the dimension and the entropy of the shared random variable in a local model.

\end{abstract}

\maketitle

Bell inequalities limit the strength of possible correlations for any model satisfying a natural definition of locality formulated by Bell~\cite{bell64}. Thus, the violation of a Bell inequality indicates that no local model (in the sense of Bell) can reproduce the observed data, which is therefore said to be (Bell) nonlocal. This can be the case in quantum theory where, by performing well-chosen local measurements on an entangled quantum system, one may obtain nonlocal quantum correlations. 

Beyond their fundamental interest, Bell inequalities are widely used tools, in particular in quantum information science~\cite{review}. Indeed, a Bell inequality violation represents a simple and strong test for certifying the presence of entanglement. Importantly this test is device-independent, in the sense that, assuming quantum theory, a Bell inequality violation implies the presence of entanglement without any assumption on the measuring devices, nor on the Hilbert space dimension. Moreover, the violation of a Bell inequality can also be used to certify the presence of genuine quantum randomness~\cite{colbeck,pironio}, or the security of a cryptographic key~\cite{acin}.

Consider two separate observers, performing local measurement on a shared physical system. The experiment results in some data, namely the joint conditional probabilities of observing a pair of outputs (measurement results) given a pair of inputs (choice of measurement settings). As they capture the strength of correlations in the data, it is intuitive that Bell inequalities are constructed based on some measure of correlation. For instance, the simplest and most famous Bell inequality of Clauser-Horne-Shimony-Holt (CHSH)~\cite{chsh} is a linear combination of simple correlation functions (see below). As these correlation functions are themselves linear combinations of the joint conditional probabilities, the CHSH inequality is a so-called linear Bell inequality. More generally, it is known that the set of distributions that are Bell local can be fully characterized via linear Bell inequalities~\cite{Pitowsky89}, and great efforts have been dedicated to find such inequalities, see e.g. Refs.~\cite{pitowsky,sliwa,CG,I4422,tamas,rosset}. 

It is however natural to ask if other measures of correlations can be used for devising Bell inequalities. Beyond the purely conceptual interest, this could be relevant in a practical context in which the full data is not available, but only certain specific (not necessarily linear) functions of the joint probabilities are. Moreover, it would be interesting to devise new types of Bell inequalities for the case of continuous measurement outcomes, for which the standard approach for constructing linear inequalities does not work anymore. 

Several works have explored these ideas. First, entropic Bell inequalities~\cite{bc,cerf,chaves} were obtained by considering the mutual information between the measurement outcomes. Later, inequalities based on higher moments of the distribution were also derived~\cite{eric,salles,belzig,bednorz}. Finally, nonlinear Bell inequalities have been developed for discussing the generalization of Bell nonlocality to networks~\cite{branciard,biloc2,armin,chaves16,rosset16}.

Here we introduce a novel class of nonlinear Bell inequalities based on the covariance of measurement results, a natural measure of correlations that is widely used in many different areas of science. We propose a general way to construct such Bell inequalities for covariances, and present, in the simplest case, a method for computing their local bound---a nontrivial problem due to the nonlinearity of the Bell expression. We show on a few explicit examples that our inequalities can detect quantum nonlocality. Moreover, we show that covariance Bell inequalities can be used to characterize shared randomness in a Bell test, providing device-independent lower bounds on the dimension and the entropy of the shared random variable in any local model.

\paragraph{Standard Bell inequalities.---}

Consider an experiment in which two parties, Alice and Bob, can both locally test different properties of a shared physical system. Let us label by $x$ the measurement choice of Alice and $y$ that of Bob, with the corresponding outcomes defining the random variables $A_x$ and $B_y$, which take the values $a$ and $b$, respectively. We will only consider non-signalling scenarios, in which the statistics of $A_x$ do not depend on the choice of $y$, and similarly the statistics of $B_y$ do not depend on $x$.

The statistics of the whole experiment, and thereby the correlations between Alice and Bob's measurement outcomes, are characterized by the joint conditional probability distribution $P(a,b|x,y)=P(A_x=a, B_y=b)$. Such a distribution is called local (in the sense of Bell) if it admits a decomposition of the form  
\ba
\label{eq:local} P(a,b|x,y) &= \int \rho(\lambda) \; P_A (a|x,\lambda ) \; P_B (b|y,\lambda) \; d\lambda ,
\ea
where $\lambda$ represents the possible values a shared classical variable $\Lambda$ (shared randomness), distributed with the density function $\rho(\lambda)$. The local response functions, defined by the distributions $P_A (a|x,\lambda )$ and $P_B (b|y,\lambda)$ represent the local behavior of Alice and Bob's subsystems.

The set of local distributions---the ``local set''---is constrained by linear Bell inequalities of the form
\ba
 \text{B} \,= \sum_{x,y,a,b} \alpha_{ab|xy} \ P(a,b|x,y) \,\leq\, \beta,
\ea
with some real coefficients $\alpha_{ab|xy}$, and where the ``local bound'' $\beta$ is the maximal value of the quantity B over all local distributions. For any given (finite) number of possible measurement settings and outcomes, the local set forms a polytope, and is thus fully characterized by a finite set of such linear Bell inequalities~\cite{Pitowsky89}.

As a concrete example, consider the case of two measurements for each party, labelled by $x,y \in \{0,1\}$, with binary outcomes $a,b \in \{+1,-1 \}$. Here we have the well-known CHSH inequality~\cite{chsh} 
\ba \label{eq:chsh}
\text{CHSH} \equiv \mean{A_0 B_0} {+} \mean{A_0 B_1}{+}\mean{A_1 B_0}{-} \mean{A_1 B_1} \leq 2, \quad\
\ea
where the correlation functions $\mean{A_x B_y}$ are simply defined as the expectation values of the products of outcomes, $\mean{A_x B_y} = \sum_{a,b} a\,b \, P(a,b|x,y)$. 
Like many other linear Bell inequalities, the CHSH inequality can detect quantum nonlocal correlations, obtained by performing well chosen local measurements on a shared entangled state.

\paragraph{A Bell inequality for covariances.---}

Instead of using the correlation functions $\mean{A_x B_y}$, other quantities can also be considered to characterize local distributions; indeed, Bell inequalities were for instance constructed for entropic quantities~\cite{bc,cerf,chaves}.
In this work we focus on another natural and widely used measure of correlation, namely the covariance. As mentioned previously, this may be of practical interest for situations where the full data is not available, and only covariances can be estimated. An advantage of using covariances is also that, like entropies, they are naturally defined for any number of possible measurement outcomes, and even for continuous outcomes. Unlike entropies however, they depend on the specific values given to the measurement outcomes; this allows one, in particular, to distinguish correlations versus anti-correlations.

The covariance of the two output variables $A_x$ and $B_y$ is defined as
\ba \label{eq:def_cov}
\text{cov}(A_x,B_y) = \mean{A_x B_y} - \mean{A_x} \mean{ B_y},
\ea
with $\mean{A_x B_y}$ defined as above, and similarly with $\mean{A_x} = \sum_{a,b} a \, P(a,b|x,y)$ and $\mean{B_y} = \sum_{a,b} b \, P(a,b|x,y)$ (for continuous values, the sums can simply be replaced by integrals, and the probabilities by probability density functions; note also that because of the non-signalling assumption, $\mean{A_x}$ does not depend on $y$, and $\mean{B_y}$ does not depend on $x$). We emphasize that because of the product term in Eq.~\eqref{eq:def_cov}, the covariance is a nonlinear function of the joint probabilities $P(a,b|x,y)$.

We start by again considering the case of two measurements per party ($x,y \in \{0,1\}$), but with now outcomes $A_x$ and $B_y$ that can take any values $a,b$ in the interval $[-1,+1]$; this implies in particular that $-1 \leq \text{cov}(A_x,B_y) \leq 1$ (note that if $a,b$ are taken in any other bounded interval, they can, together with the corresponding covariance, simply be rescaled so that they lie in $[-1,+1]$).
Our goal is to bound the set of possible values of the four covariances $\text{cov}(A_x,B_y)$ for local distributions---the ``local set'' for covariances. To get some intuition, and inspired by the form of CHSH in Eq.~\eqref{eq:chsh}, one may look for instance at the projection of this local set onto the 2-dimensional $(\text{covCHSH},\text{covCHSH}')$ plane, with
\ba  \label{eq:covCHSH}
\text{covCHSH} &\equiv& \text{cov}(A_0,B_0) +\text{cov}(A_0,B_1) \nonumber \\ & & + \ \text{cov}(A_1,B_0) -\text{cov}(A_1,B_1) , 
\ea
and with $\text{covCHSH}'$ defined similarly, except that $\text{cov}(A_0,B_1)$ and $\text{cov}(A_1,B_0)$ come with minus signs.
\begin{figure}
	\begin{center}
	\includegraphics[width=.95\columnwidth]{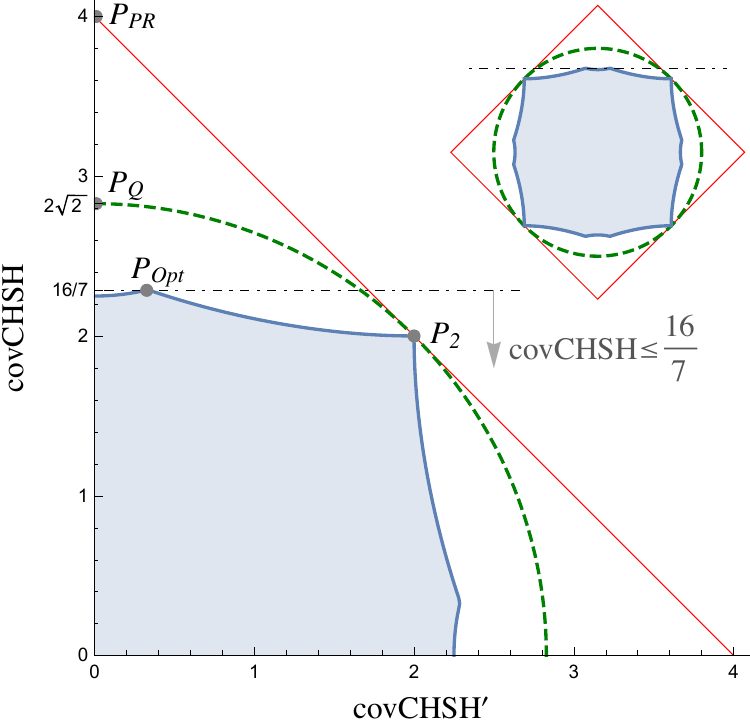}
	\end{center}
	\caption{Projection of the local set for covariances (shaded region) onto the upper-right quadrant of the $(\text{covCHSH},\text{covCHSH}')$ plane; the other three quadrants are obtained by symmetry, see inset. All local distributions give values in the local set and satisfy the covariance Bell inequality $\text{covCHSH} \le \frac{16}{7}$. This inequality is saturated by the distributions that project onto $P_{Opt}$; those require mixing at least 3 deterministic strategies, while mixing 2 allows one to obtain for instance the point $P_2$ (see main text). Quantum distributions can reach the dashed green circle and can thus violate the covariance Bell inequality, with $P_Q$ corresponding to a distribution that also maximally violates the CHSH inequality. The no-signaling set is limited by the solid red line (as it satisfies $\text{covCHSH} + \text{covCHSH}' = 2[\text{cov}(A_0,B_0)-\text{cov}(A_1,B_1)] \le 4$), where $P_{PR}$ corresponds to a PR nonlocal box.}
	\label{fig:geom}
\end{figure}
By optimizing the values of $\text{covCHSH}^{(\prime)}$ numerically (but with reliable enough results), we obtained the local set represented in Fig.~\ref{fig:geom}.

One immediately sees that contrary to the standard local polytope in the probability space, the local set in the covariance space is nonconvex, and that (unsurprisingly) it cannot be fully described by a finite number of Bell inequalities based on linear combinations of the covariances. Its full characterization thus looks much more complicated than that of the standard local polytope. Nevertheless, one can still derive some Bell inequalities for covariances that provide an outer approximation of the local set. As illustrated on Fig.~\ref{fig:geom}, an example of such covariance Bell inequalities is the following:
\ba  \label{ineq:covCHSH}
\text{covCHSH} \, \leq \, \frac{16}{7} \, .
\ea

In order to prove that this inequality indeed holds for any local distribution, it is in fact sufficient to restrict to distributions with binary outputs $\pm 1$. Those can be written as convex mixtures of finitely many deterministic (local) distributions, and one can then optimize the expression $\text{covCHSH}$ over the corresponding weights. To take into account the constraint that these weights must be nonnegative and sum up to $1$, one can introduce Karush-Kuhn-Tucker (KKT) multipliers~\cite{KKT}, which provide necessary conditions for a solution to be optimal. By considering decompositions onto different sets of deterministic distributions with nonzero weights, these KKT conditions simplify to a number of linear systems, which can easily be solved. The full details of the proof are given (together with an alternative approach) in Appendix~\ref{app:proofs}.

It is instructive to look more closely in the proof at decompositions with a given number $d$ of deterministic distributions with nonzero weights. First note that for $d=1$, i.e., for deterministic distributions, all covariances are $0$, and therefore one also obtains $\text{covCHSH}=0$. To obtain a nonzero value, one thus needs to consider mixtures of at least 2 distributions. For $d=2$, one finds that the maximal possible value is $\text{covCHSH}=2$, obtained for instance by the distribution \mbox{$P_2 = \frac12 \big[(++/+-) + (--/-+)\big]$} (which also gives $\text{covCHSH}'=2$, see Fig.~\ref{fig:geom})---where the notation $(A_0A_1/B_0B_1)$ denotes a strategy in which Alice and Bob deterministically output $A_x$ and $B_y$ for inputs $x$ and $y$, respectively. In order to reach the local upper bound in Eq.~\eqref{ineq:covCHSH}, one needs to go to $d=3$; the distribution $P_{Opt} = \frac37(++/++)+\frac27(-+/--)+\frac27(--/-+)$, for instance, gives $\text{covCHSH}=\frac{16}{7}$ (and $\text{covCHSH}'=\frac{16}{49}$, see Fig.~\ref{fig:geom}). One thus sees that reaching a given value of $\text{covCHSH}$ requires mixing a certain number of deterministic strategies; this is an interesting feature, which will allow one to use the covariance inequality as a shared randomness witness (see below).

One may also look at possible violations of the covariance Bell inequality~\eqref{ineq:covCHSH}. One finds that it can indeed detect quantum nonlocality: that is, one can obtain $\text{covCHSH} > \frac{16}{7}$ by performing local measurements on an entangled state.
For example, the quantum correlations that reach the maximal quantum value (the ``Tsirelson bound''~\cite{Tsirelson}) of CHSH=$2\sqrt{2}$ have vanishing marginals $\mean{A_x} = \mean{ B_y}= 0$, and therefore also give covCHSH=$2\sqrt{2} > \frac{16}{7}$ (point $P_Q$ on Fig.~\ref{fig:geom}). In fact, this is the largest possible quantum violation, as one can prove that all quantum correlations must satisfy the inequality
\begin{equation} \label{ineq:covCHSH_tsirelson}
\text{covCHSH} \leq 2 \sqrt{2} \,.
\end{equation}
This is shown in Appendix~\ref{app:Tsirelson_bnd}, where we also give a tighter characterization of the quantum set in terms of covariances. Note that in contrast to the local bound, marginals do not play any role in reaching the Tsirelson bound for covCHSH. Nevertheless, the nonlinearity of the Bell expression $\text{covCHSH}$ also has interesting consequences in the quantum case. For instance, one can find two pure entangled states such that none of the states can individually violate the inequality~\eqref{ineq:covCHSH}, whereas a mixture of the two states can violate it; see Appendix~\ref{app:quantum_mixed} for details.

Note, finally, that given the range of output variables, we have $|\text{cov}(A_x,B_y)| \leq 1$, and therefore the algebraic maximum possible value of $\text{covCHSH}$ is $4$ (just like for CHSH). This value can be reached by the non-signalling distribution known as the Popescu-Rohrlich (PR) nonlocal box~\cite{PR} (point $P_{PR}$ on Fig.~\ref{fig:geom}), which again has vanishing marginals and also reaches the algebraic maximum of the CHSH expression.

\paragraph{Constructing other covariance Bell inequalities.---} One can follow similar ideas to the ones developed above and derive other Bell inequalities based on covariances. 

In the case of ternary inputs $x,y \in \{0,1,2\}$ and binary outputs $a,b \in \{+1,-1\}$ for instance, the local set in the space of expectation values $\mean{A_x B_y}$ is again fully characterized by CHSH-like inequalities; however, when considering the full probability space, i.e., including the marginals $\mean{A_x}$, $\mean{B_y}$, one gets a new inequivalent family of Bell inequalities of the form~\cite{froissart,sliwa,CG}
\ba
I_{3322} &\equiv & \mean{A_0B_0} + \mean{A_0B_1} + \mean{A_0B_2} + \mean{A_1B_0} \nonumber \\
&& \ + \mean{A_1B_1} - \mean{A_1B_2} + \mean{A_2B_0} - \mean{A_2B_1} \nonumber \\
&& \quad + \mean{A_0} + \mean{A_1} - \mean{B_0} - \mean{B_1} \ \leq \ 4. \quad \label{ineq:I3322}
\ea

Inspired by the previous example of CHSH, one may look at the local bound when the expectation values $\mean{A_x B_y}$ are replaced by the covariances $\text{cov}(A_x,B_y)$. Marginal terms, viewed e.g. as $\mean{A_x} = \mean{A_x \openone_B}$ (where $\openone_B$ is the identity measurement operator for Bob, that always outputs $b=1$) then simply drop out, as $\text{cov}(A_x,\openone_B)=0$.
We thus get the covariance Bell inequality (that allows for any outputs $a,b \in [-1,+1]$)
\ba  
\text{cov3322} &\equiv& \text{cov}(A_0,B_0) +\text{cov}(A_0,B_1) +\text{cov}(A_0,B_2) \nonumber \\ 
& & \ + \ \text{cov}(A_1,B_0) +\text{cov}(A_1,B_1)-\text{cov}(A_1,B_2) \nonumber \\
& & \quad + \ \text{cov}(A_2,B_0)-\text{cov}(A_2,B_1) \ \leq \ \frac{9}{2} \, . \label{ineq:cov3322}
\ea
Here the local bound was obtained through numerical optimization (up to machine precision); nevertheless, the reproducibility and reliability of the numerical results make us quite confident that the bound is correct.
In Appendix~\ref{app:local_strats} we present local distributions that reach it, and discuss quantum and super-quantum (non-signalling) violations.

More generally, one may consider constructing covariance Bell inequalities following the above recipe, starting from an arbitrary linear Bell inequality with binary outcomes. Note that instead of simply dropping the marginal terms, one could as well keep them, or replace them with other functions, e.g. with variances.

Generalising even further our approach for covariance Bell inequalities, one can also investigate Bell inequalities based on Pearson correlators, which can be thought of as normalised covariances. The Pearson correlator for two variables $A_x$ and $B_y$ with variances $\sigma(A_x)^2 = \mean{A_x^2} - \mean{A_x}^2$ and $\sigma(B_y)^2 = \mean{B_y^2} - \mean{B_y}^2$ is defined as $\text{r}(A_x,B_y) = \frac{ \text{cov}(A_x, B_y) } { \sigma(A_x) \sigma(B_y)}$ (if either $\sigma(A_x)$ or $\sigma(B_y)$ is zero, we define $\text{r}(A_x,B_y) = 0$). Following a similar construction as above, we get, for the case of binary inputs and now also restricting to binary outputs, the Bell inequality
\ba  \label{ineq:rCHSH}
\text{rCHSH} &\equiv & \text{r}(A_0,B_0) +\text{r}(A_0,B_1) \nonumber \\[-1mm] & & 
+ \ \text{r}(A_1,B_0) -\text{r}(A_1,B_1) \ \leq \ \frac{5}{2} ,
\ea
where the local bound was again obtained through numerical optimization.
It must be emphasized that unlike for covariances, the local set for Pearson correlators is not the same when considering binary outputs or more possible outputs; for instance, with ternary outputs in $\{+1,0,-1\}$, one can locally reach the value $\text{rCHSH} = 2\sqrt{2}$, which turns out to also be the Tsirelson bound for rCHSH (whether we restrict to binary outputs or not), as proven already in Ref.~\cite{landau} (we also give a proof of this in Appendix~\ref{app:Tsirelson_bnd}).
In Appendix~\ref{app:local_strats} we present local distributions that reach these two local bounds of $\frac{5}{2}$ and $2\sqrt{2}$, and discuss quantum and super-quantum violations of them.
Of course, similar Bell inequalities with Pearson correlators could also be constructed, following a similar recipe as suggested above, starting e.g. from $I_{3322}$ or any other linear Bell inequality with binary outcomes.

\paragraph{Shared randomness witnesses.---}

Let us come back now to covCHSH. As emphasized before, the value of covCHSH that can be obtained locally depends on the number of deterministic strategies involved in the local strategy being used. This readily allows one to obtain a lower bound on the classical dimension $d$ of the shared variable $\Lambda$ (i.e., the number of different values it can take), or equivalently its max-entropy $H_{\text{max}}(\Lambda) \equiv \log_2 d$: from the discussion above, it follows that as soon as $\text{covCHSH} > 0$, one requires $d \ge 2$, i.e. $H_{\text{max}}(\Lambda) \ge 1$; if $\text{covCHSH} > 2$, then $d \ge 3$, i.e. $H_{\text{max}}(\Lambda) \ge \log_2 3$. 

\begin{figure}
	\begin{center}
	\includegraphics[width=.86\columnwidth]{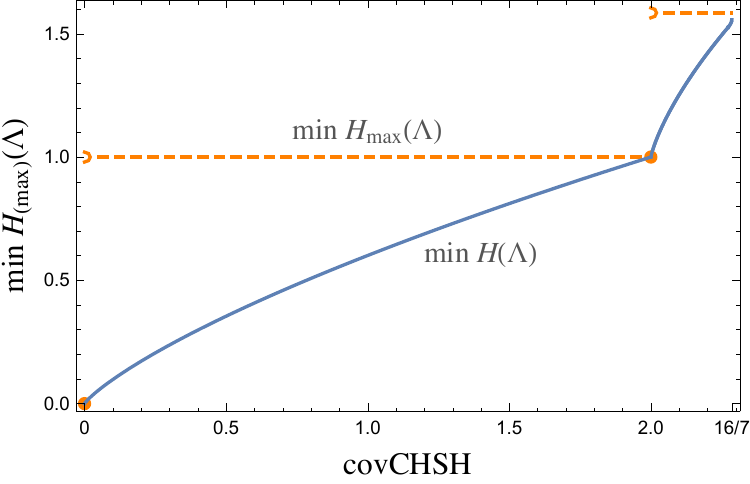}
	\end{center}
	\caption{Minimum Shannon entropy $H(\Lambda)$ (blue curve) and max-entropy $H_{\text{max}}(\Lambda) = \log_2 d$ (orange dashed curve) of the shared variable $\Lambda$ that is required to obtain any given value of the Bell expression covCHSH with a local model. For instance, obtaining $\text{covCHSH}=2$ requires at least one bit of (Shannon or max-) entropy, while reaching the maximum local value $\text{covCHSH}=\frac{16}{7} \simeq 2.29$ requires $H(\Lambda) \gtrsim 1.56$ and $H_{\text{max}}(\Lambda) \geq \log_2 3 \simeq 1.58$, i.e. $d \geq 3$.}
	\label{fig:entropy}
\end{figure}

One may also quantify the amount of shared randomness in terms of the Shannon entropy of $\Lambda$, $H(\Lambda) \equiv - \sum_{\lambda} q_\lambda \log_2 q_\lambda$ (defined here, for simplicity, for the case of a discrete variable $\Lambda$ that takes the value $\lambda$ with probability $P(\Lambda=\lambda) = q_\lambda$). For a given value of covCHSH between $0$ and $\frac{16}{7}$, one can also obtain a lower bound on $H(\Lambda)$ by minimizing it for all local strategies. As it turns out, it suffices to optimise $H(\Lambda)$ over decompositions onto deterministic strategies with binary outputs $\pm 1$---intuitively, local randomness does not help increase covCHSH. The details of our optimization are given in Appendix~\ref{app:min_shared_randomness}, and the results we obtained are plotted on Fig.~\ref{fig:entropy}. Unsurprisingly, we find that $\min H(\Lambda)$ increases with the value of covCHSH. As soon as $\text{covCHSH} > 0$, one requires $H(\Lambda) > 0$; for $\text{covCHSH} = 2$, one must have $H(\Lambda) \ge 1$ (recall that $\text{covCHSH} = 2$ can be reached by mixing two deterministic strategies with equal weights); finally, reaching the local bound of $\text{covCHSH} = \frac{16}{7}$ requires at least as much shared randomness as in the distribution $P_{Opt}$ given previously, i.e. $H(\Lambda) \ge -\frac{3}{7} \log_2 \frac{3}{7} -2 \!\times\! \frac{2}{7} \log_2 \frac{2}{7} \simeq 1.56$.

Thus covCHSH acts as a device-independent witness for characterizing the dimension and the entropy of the shared randomness of a local strategy in a Bell experiment. This complements recent works on device-independent tests of the dimension of quantum systems~\cite{dimH}, and of the dimension~\cite{gallego} and entropy~\cite{CBB} of classical communications.

\paragraph{Discussion.---}

In this Letter we introduced Bell inequalities based on covariances, a natural measure of correlations, rather than on linear combinations of probabilities.
We presented the simplest example of such a covariance Bell inequality with binary inputs, that echoes the well-known CHSH inequality, and investigated some of its properties. Proving analytically the local bound is not straightforward here, due to the nonlinearity of the Bell expression, but we could nevertheless provide a method to do so. Its quantum bound was also proven analytically.

Beyond this simplest example, we proposed a general recipe to construct, from any standard Bell inequality with binary outcomes, new Bell inequalities for covariances as well as for Pearson correlators. While our analytical method for proving the local bound generalizes in principle to any covariance Bell inequality obtained with our construction, in practice it becomes intractable for larger numbers of inputs. Nevertheless, for the examples we presented the local bounds could reliably be obtained numerically. It would be interesting to find a more efficient way to prove local bounds, as well as quantum bounds, for covariance-based Bell inequalities---or even to find other ways to characterize the local and quantum sets of admissible covariances (note that a tighter characterization of the quantum set than Eq.~\eqref{ineq:covCHSH_tsirelson} can already be given, see Appendix~\ref{app:Tsirelson_bnd}). One possible idea would be to look for some hierarchy of criteria---similar in spirit to that of Navascu\'es-Pironio-Ac\'in (NPA) for quantum correlations---that covariances must satisfy; indeed, one finds that covariances and Pearson correlators appear naturally in the NPA hierarchy~\cite{NPA}.

An interesting feature of our covariance Bell inequalities is that they serve as a shared randomness witness. Indeed, in contrast with standard linear Bell inequalities where the local bound can always be attained by a single deterministic strategy, reaching the local bound---or in fact, any nontrivial value---for a Bell expression defined in terms of covariances requires, for a local strategy, to make use of shared randomness. We showed explicitly, on our simplest example, how the value of the Bell expression allows one to place device-independent lower bounds on the amount of shared randomness in a local model (both in terms of its dimension and its entropy).
Such witnesses may help addressing certain problems in quantum nonlocality related to shared randomness~\cite{pal,sikora,vicente}, in particular finding what is the minimal amount of shared randomness necessary to simulate the correlations of entangled states admitting a local model~\cite{joe}. 
Our example with binary inputs allows one to certify the use of relatively little shared randomness (e.g. a dimension of at least $3$); it would be interesting to find (families of) covariance Bell inequalities that can certify larger amounts of shared randomness. While covariance Bell inequalities were found to provide a natural way to provide device-independent witnesses for shared randomness, such witnesses could also be studied in other frameworks, with other types of Bell inequalities that require shared randomness to be saturated---e.g. with entropic inequalities~\cite{bc,cerf,chaves}.

From a more practical perspective, our covariance Bell inequality could be useful in experimental situations where access to measurement data is limited, e.g. if only covariances can be measured, and where standard Bell inequalities cannot be used. This could be relevant to certain mesoscopic electronics experimental setups.
With such applications in mind, it would also be interesting to also investigate similar covariance inequalities to the ones constructed here for other types of quantum nonlocality, e.g. for entanglement~\cite{HHHH} or quantum steering~\cite{steering} (we note that inequalities were constructed in these contexts for covariance matrices, although following very different approaches~\cite{WW,giedke,guehne,jevtic,maccone}).

Finally, Bell inequalities based on covariances could be useful for the study of nonlocal correlations in networks. When a network features several sources that are assumed to be independent, the set of local distributions (or rather, ``$N$-local distributions'' for a network with $N$ independent sources) is typically nonconvex, and nonlinear Bell inequalities are necessary to give an effective description of the ($N$-)local set~\cite{branciard,biloc2,armin,chaves16,rosset16}. In fact, some recent work~\cite{chaves17} has already made use of covariances in the context of quantum networks, however not constructing explicit Bell inequalities. The ideas developed here could also find some nice applications in providing ways to obtain new types of Bell inequalities, and possibly shared randomness witnesses, for networks.

\paragraph{Acknowledgements.---} We thank Alastair A. Abbott, Joe Bowles, Michael J. W. Hall and Paul Skrzypczyk for discussions. We acknowledge financial support from the Swiss National Science Foundation (Starting grant DIAQ and QSIT) and from the French National Research Agency (`Retour Post-Doctorants' program ANR-13-PDOC-0026).



\clearpage

\appendix
\setcounter{secnumdepth}{3}

\section{Proofs of the covariance Bell inequality~\eqref{ineq:covCHSH}}
\label{app:proofs}

Obtaining the local bound for a standard (linear) Bell inequality is rather straightforward, as one can simply focus on deterministic strategies. However, such an approach does not work for covariance Bell inequalities because of their nonlinearity.

In this Appendix we give the details of the proof of the inequality~\eqref{ineq:covCHSH}, as sketched in the main text. We then present an alternative proof that also uses Karush-Kuhn-Tucker (KKT) multipliers and conditions~\cite{KKT}, although in a different way.
While the former can in principle be adapted to any other covariance Bell inequality (but may not actually be practical because of the very high number of cases to consider for increased numbers of parameters), the latter exploits specific properties of the CHSH (2-input / 2-output) scenario and uses fewer variables in the optimization; it is however less instructive with respect to the shared randomness required for obtaining a given value of covCHSH with a local model, and adapting it to another scenario will first require finding the right properties to exploit.

\subsection{Optimizing over the weights of deterministic distributions}
\label{app:proof_v1}

The proof here will be divided into three parts. The first one consists in showing that, without loss of generality, we can focus on a situation where only binary outputs are considered. The second part describes how standard methods for solving (quadratic) constrained problems can be applied. Finally, the third part shows how our quadratic problem can be simplified to a number of linear ones, and solved.

\subsubsection{Reduction to binary outputs}

Consider a local distribution $P(a,b|x,y)$ with outputs $a,b \in [-1,+1]$.%
\footnote{Note that given the translational invariance of covariances (i.e., $\text{cov}(A_x,B_y) = \text{cov}(A_x+\alpha,B_y+\beta)$ for any real values $\alpha, \beta$), it would be equivalent to take $a,b$ in any interval of length $2$. Similarly in the case of binary outputs, instead of $\pm 1$ one could take any two real values with a difference of $2$.}
Let then Alice and Bob post-process their outputs locally, in the following way: they replace each output $a$ ($b$) by $a'=+1$ ($b'=+1$) with probability $\frac{1+a}{2}$ ($\frac{1+b}{2}$), or by $a'=-1$ ($b'=-1$) with probability $\frac{1-a}{2}$ ($\frac{1-b}{2}$), thus defining a new local distribution $P'(a',b'|x,y) = \sum_{ab} \frac{1+a' a}{2} \frac{1+b' b}{2} P(a,b|x,y)$ with binary outputs $a',b' \in \{+1,-1\}$. One can easily check that the expectation values $\mean{A_x^{(\prime)}B_y^{(\prime)}}$, $\mean{A_x^{(\prime)}}$ and $\mean{B_y^{(\prime)}}$, and therefore also the covariances $\text{cov}(A_x^{(\prime)},B_y^{(\prime)})$, are the same for the two local distributions $P$ and $P'$.%
\footnote{Note, however, that the Pearson correlators may in general be different, with $|\text{r}(A_x',B_y')| \le |\text{r}(A_x,B_y)|$. This is why one does not obtain the same local bound on rCHSH when considering binary outputs or a larger set of possible outputs; see Appendix~\ref{app:rCHSH}.}

Thus, any values of the covariances $\text{cov}(A_x,B_y)$ (for the various inputs $x,y$) obtained by a local distribution with inputs in $[-1,+1]$ can also be obtained by a local distribution with inputs in $\{+1,-1\}$ (and vice-versa, obviously). The local set for covariances is the same in both cases, and to characterize it it thus suffices to restrict to local distributions with binary outputs $\pm1$.

\subsubsection{KKT conditions}

Our goal is now to find the largest possible value of the quantity covCHSH for any local strategy with binary inputs and outputs. Such a local distribution can be represented by a 16-dimensional vector $\vec{P}$, the components of which are the 16 joint probabilities $P(a,b|x,y)$\footnote{For simplicity we may identify a distribution $P$ with its vectorial representation $\vec P$.}---and which in fact, because of the normalization and non-signaling constraints, live in a space of dimension only $\text{dim}_{\vec P} = 8$.%
\footnote{An alternative, more compact parametrization for $\vec{P}$ may for instance be given by the 8 correlators $\mean{A_xB_y}$, $\mean{A_x}$ and $\mean{B_y}$; see Appendix~\ref{app:proof_v2}.}
Moreover, $\vec{P}$ admits a convex decomposition in terms of the vertices of the local polytope of the form $\vec{P} = \sum_k q_k \vec{P}_k^{\text{det}}$ with $q_k \geq 0$ , $\sum_k q_k = 1$, and where the vectors $\vec{P}_k^{\text{det}}$ represent the 16 deterministic local strategies. (Note already that due to Carath\'eodory's theorem~\cite{caratheodory}, for any local distribution $\vec{P}$ it will in fact be sufficient to consider at most $\text{dim}_{\vec P}+1=9$ deterministic strategies in the decomposition.)
Evaluating covCHSH, we get
\ba \label{eq:covCHSH_decomp}
\text{covCHSH}(\{q_k\}_k) = \sum_k q_k C_{kk} - \sum_{i,j} q_i q_j C_{ij} , \quad
\ea
where $C_{ij} = A_0^i B_0^j + A_0^i B_1^j + A_1^i B_0^j - A_1^i B_1^j = \pm 2$ and $A_x^i$ refers to Alice's output for the measurement $x$ given that strategy $\vec{P}_i^{\text{det}}$ is used, and similarly for $B_y^j$---so that in particular, $C_{kk} = \text{CHSH}(\vec{P}_k^{\text{det}})$ is the CHSH value obtained for the deterministic strategy $\vec{P}_k^{\text{det}}$.

We shall now maximize $\text{covCHSH}(\{q_k\}_k)$ with respect to the variables $q_k$---which, because of the second term in Eq.~\eqref{eq:covCHSH_decomp}, is a nonlinear optimization problem. To account for the constraints that $q_k \geq 0 $ and $\sum_k q_k = 1$, we introduce KKT multipliers (i.e., Lagrange-like multipliers) $\lambda_k$ and $\mu$, and define the following Lagrangian:
\ba \label{eq:lagrangian_v1}
L & = & \text{covCHSH}(\{q_k\}_k) + \sum_k \lambda_k \, q_k + \mu\,(1 {-} \sum_k q_k) \nonumber \\[-1mm]
& = & \sum_k q_k C_k - \sum_{i,j} q_i q_j Z_{ij} + \sum_k \lambda_k \, q_k + \mu\,(1 {-} \sum_k q_k) . \nonumber \\[-2mm]
\ea
Given the regularity of the problem, with the constraints being affine functions of the variables, the (local) maxima must then necessarily fulfill the KKT conditions~\cite{KKT}: 
\ba \label{eq:KKTcond_v1}
& & \frac{\partial L}{\partial q_k} = C_{kk} - \sum_i q_i (C_{ik} + C_{ki} ) + \lambda_k - \mu = 0  \quad \forall k, \nonumber \\ 
& & \frac{\partial L}{\partial \mu} = 1 - \sum_k q_k = 0 , \nonumber \\
& & \lambda_k q_k = 0  , \; \lambda_k \geq 0 , \;  q_k \geq 0  .
\ea

\subsubsection{Reduction to linear systems of equations and resolution}

Because of the constraint $\lambda_k q_k = 0$, the above system of equations is nonlinear. One can however reduce it to a number of different linear systems of equations by imposing that a given subset of weigths $q_k$ take nonzero values, and by considering separately the case of each possible subset. For those $q_k$ in the given subset under consideration, the last line of Eq.~\eqref{eq:KKTcond_v1} simply becomes $\lambda_k = 0, q_k > 0$, while for the other $q_k$'s, it becomes $q_k = 0 , \lambda_k \geq 0$---with in both cases, $\lambda_k = - C_{kk} + \sum_i q_i (C_{ik} + C_{ki} ) + \mu$ according to the first line of Eq.~\eqref{eq:KKTcond_v1}.
For a given number $d$ of nonzero weights, the system~\eqref{eq:KKTcond_v1} then reduces to $d+1$ linear equality constraints and $16$ (also linear) inequality constraints (with $d$ strict and $16-d$ nonstrict inequalities) for the remaining $d+1$ nontrivial variables $q_k$ and $\mu$.

Note that if only $d=1$ weight $q_k$ is nonzero (and therefore equal to 1), then the distribution $P(a,b|x,y)$ is deterministic, which implies that all covariances---and hence covCHSH---are zero.
Furthermore, as already mentioned previously, due to Carath\'eodory's theorem it is sufficient to only consider subsets with at most $d=9$ (out of $16$) nonzero $q_k$'s---indeed, if a maximum is reached by some decomposition involving more than $9$ deterministic distributions, then it is also reached by another decomposition giving the same distribution, but involving no more than $9$ deterministic distributions.
We used Mathematica to solve, for each value of $d$ between $2$ and $9$, the $\binom{16}{d}$ different linear systems of equations corresponding to all the different subsets containing $d$ nonzero weights (which makes a total of $\sum_{d=2}^9 \binom{16}{d} = 50\,626$ linear systems to consider). More specifically, we first found the solutions that satisfy all equality constraints and then checked which ones also satisfy the inequality constraints. To obtain the local bound of our covariance Bell inequality~\eqref{ineq:covCHSH}, there then remains to check what is the largest possible value of covCHSH for all these feasible solutions.

For $d=2$, all $\binom{16}{2} = 120$ systems of equations have solutions that satisfy all equality constraints (out of which, $56$ have unique solutions and $64$ are still underdetermined), but only $4$ out of them have (unique) solutions that also satisfy the inequality constraints. Those $4$ solutions all give $\text{covCHSH} = 2$, which is the maximal value that one can reach by mixing only $2$ local deterministic distributions.

With $d=3$ nonzero weights $q_k$, again all $\binom{16}{3} = 560$ systems of equations have solutions that satisfy all equality constraints ($432$ with unique solutions and $128$ still underdetermined), but only $8$ of them have (unique) solutions that also satisfy the inequality constraints. 
Those solutions are listed in Table~\ref{table:optim_qs}; they all have the similar form $P = \frac37 P_{k_1}^{\text{det}} + \frac27 P_{k_2}^{\text{det}} + \frac27 P_{k_3}^{\text{det}}$, and all give the value $\text{covCHSH} = \frac{16}{7}$.

\begin{table}
\centering
\begin{tabular}{ccccccccc}
 $P \ =$ & $\frac{3}{7} \ P_{k_1}^{\text{det}}$ & + & $\frac{2}{7} \ P_{k_2}^{\text{det}}$ & + & $\frac{2}{7} \ P_{k_3}^{\text{det}}$ \\[1mm]
 \hline
 \\[-4mm]
\ldelim\{{2}{3mm}[] & $\frac{3}{7} (++/++)$ & + & $\frac{2}{7} (-+/--)$ & + & $\frac{2}{7} (--/-+)$ \\[1mm]
 & $\frac{3}{7} (--/--)$ & + & $\frac{2}{7} (+-/++)$ & + & $\frac{2}{7} (++/+-)$ \\[1mm]
\ldelim\{{2}{3mm}[] & $\frac{3}{7} (+-/-+)$ & + & $\frac{2}{7} (++/+-)$ & + & $\frac{2}{7} (-+/--)$ \\[1mm]
 & $\frac{3}{7} (-+/+-)$ & + & $\frac{2}{7} (--/-+)$ & + & $\frac{2}{7} (+-/++)$ \\[1mm]
\ldelim\{{2}{3mm}[] & $\frac{3}{7} (++/+-)$ & + & $\frac{2}{7} (+-/-+)$ & + & $\frac{2}{7} (--/--)$ \\[1mm]
 & $\frac{3}{7} (--/-+)$ & + & $\frac{2}{7} (-+/+-)$ & + & $\frac{2}{7} (++/++)$ \\[1mm]
\ldelim\{{2}{3mm}[] & $\frac{3}{7} (+-/++)$ & + & $\frac{2}{7} (-+/+-)$ & + & $\frac{2}{7} (--/--)$ \\[1mm]
 & $\frac{3}{7} (-+/--)$ & + & $\frac{2}{7} (+-/-+)$ & + & $\frac{2}{7} (++/++)$
\end{tabular}
\caption{Optimal decompositions of the form $P = \frac37 P_{k_1}^{\text{det}} + \frac27 P_{k_2}^{\text{det}} + \frac27 P_{k_3}^{\text{det}}$ obtained as solutions of the KKT conditions of Eq.~\eqref{eq:KKTcond_v1}, giving $\text{covCHSH} = \frac{16}{7}$. The first 4 distributions give $\text{covCHSH}' = \frac{16}{49}$ and thus project onto the point $P_{Opt}$ on Fig.~\ref{fig:geom} (while the last 4 give $\text{covCHSH}' = -\frac{16}{49}$; note also that there exist 8 other similar decompositions that give $\text{covCHSH} = -\frac{16}{7}$ and $\text{covCHSH}' = \pm\frac{16}{49}$). The distributions in the table are grouped by pairs (brackets on the left), in which the two decompositions are obtained from one another by flipping all outputs.}
\label{table:optim_qs}
\end{table}

As it turns out, this value of $\frac{16}{7}$ obtained by suitably mixing 3 deterministic local distributions cannot be increased by mixing more distributions. For $d=4$ for instance, we find that out of the $\binom{16}{4} = 1\,820$ systems of equations, only $1\,516$ have solutions that satisfy all equality constraints, out of which only $14$ have solutions that also satisfy the inequality constraints. $6$ of these (still underdetermined) solutions give a local maximum of $\text{covCHSH} = 2$, while the other $8$ (unique) solutions---all of the form $P = \frac38 P_{k_1}^{\text{det}} + \frac38 P_{k_2}^{\text{det}} + \frac18 P_{k_3}^{\text{det}} + \frac18 P_{k_4}^{\text{det}}$ (one solution being obtained for instance by taking $P_{k_1}^{\text{det}} = (++/++), P_{k_2}^{\text{det}} = (-+/--), P_{k_3}^{\text{det}} = (+-/-+)$ and $P_{k_4}^{\text{det}} = (--/-+)$)---give $\text{covCHSH} = \frac{9}{4}$.
The results for all values of $d$ from $2$ to $9$ are summarized in Table~\ref{table:summary_solutions}. For $d=6$ and $8$, we also get local maxima of $\text{covCHSH} = 2$, while for $d=5,7$ and $9$, no solutions are found for any of the corresponding systems of KKT conditions.
As one can see, the maximal value of covCHSH obtained for all cases is thus $\frac{16}{7}$; this concludes the proof that this is indeed the value of the local bound in our covariance Bell inequality~\eqref{ineq:covCHSH}.

\begin{table}
\centering
\begin{tabular}{ccccc}
 $\quad d \quad$ & $\quad \binom{16}{d} \quad$ & {\scriptsize \begin{tabular}{c} \# consistent \\[-1mm] systems of \\[-1mm] equalities \end{tabular}} & $\ $ {\scriptsize \begin{tabular}{c} \# consistent \\[-1mm] systems incl. \\[-1mm] inequalities \end{tabular}} $\ $ & {\footnotesize \begin{tabular}{c} local max \\ covCHSH \end{tabular}} \\[3mm]
 \hline
 \\[-4mm]
2 & 120 & 120 & 4 & 2 \\[.5mm]
3 & 560 & 560 & 8 & $\frac{16}{7}$ \\[.5mm]
4 & 1\,820 & 1\,516 & 14 & $2 \ / \ \frac{9}{4}$ \\[.5mm]
5 & 4\,368 & 3\,376 & 0 & -- \\[.5mm]
6 & 8\,008 & 1\,896 & 4 & 2 \\[.5mm]
7 & 11\,440 & 688 & 0 & -- \\[.5mm]
8 & 12\,870 & 154 & 1 & 2 \\[.5mm]
9 & 11\,440 & 16 & 0 & --
\end{tabular}
\caption{For any number $d$ between $2$ and $9$ (1st column), the 2nd column of the table gives the number of subsets of deterministic distributions $P_k$ with nonzero weigths $q_k$, which defines the number of linear systems---obtained from the KKT conditions~\eqref{eq:KKTcond_v1}---to consider in the proof. The numbers of systems which have solutions when only considering the equality constraints, or when also including the inequality constraints, are given in the next two columns. The last column gives the local maxima of covCHSH obtained from these feasible  solutions of the KKT conditions. }
\label{table:summary_solutions}
\end{table}

\medskip

It should be clarified that although Table~\ref{table:summary_solutions} gives local maxima of $2$ or $\frac94$ for $d \ge 4$, this does not mean that those are the maximum values of covCHSH when mixing $d \ge 4$ deterministic local distributions. Indeed, the table gives local maxima obtained in a set delimited by strict inequalities $q_k > 0$, while the suprema over these sets may be obtained for some $q_k \to 0$. In fact, covCHSH can get arbitrarily close to $\frac{16}{7}$ with a mixture of $d \ge 4$ deterministic local distributions by mixing an optimal decomposition with $3$ distributions from Table~\ref{table:optim_qs}, with a tiny amount of $d-3$ other distributions.

Another observation of interest is that all solutions of the KKT conditions that give a local maximum value of $\text{covCHSH} = 2$ (for $d = 2,4,6,8$, as listed in the table) are of the general form $P = q_1' \big[(++/++) + (--/--)\big] + q_2' \big[(++/+-) + (--/-+)\big] + q_3' \big[(+-/++) + (-+/--)\big] + q_4' \big[(+-/-+) + (-+/+-)\big]$ with $q_1',q_2',q_3',q_4' \ge 0$ and $2(q_1'+q_2'+q_3'+q_4')=1$: for $d=2$, three of the coefficients $q_k'$ are zero (which, for each remaining nonzero $q_k'$, gives a unique solution to the corresponding system of equations); for $d=4$, two of the coefficients $q_k'$ are zero; for $d=6$, one of the coefficients $q_k'$ is zero; and for $d=8$, all coefficients $q_k'$ are nonzero (which indeed makes the solutions in these last 3 cases underdetermined).
As one can see, in this general decomposition each deterministic distribution comes with its ``opposite'', in which all outputs are flipped, with the same weight. This implies in particular that all marginal expectation values $\mean{A_x}$ and $\mean{B_y}$ are zero, that the covariances $\text{cov}(A_x,B_y)$ are equal to the expectation values $\mean{A_xB_y}$, and therefore that the expression of covCHSH simply reduces to that of CHSH (i.e., the nonlinear part vanishes)---which is indeed bounded by $2$ for local distributions.
This highlights the crucial role played by the nonlinear terms $\mean{A_x} \mean{B_y}$ in covCHSH: these are precisely the terms that allow the local bound of covCHSH to be greater than that of CHSH; and in order to reach a value of covCHSH larger than 2 with a local distribution, one needs at least one pair of settings $x,y$ for which $\mean{A_x} \mean{B_y} \neq 0$.

\subsection{Optimizing over the expectation values $\mean{A_xB_y}$, $\mean{A_x}$ and $\mean{B_y}$}
\label{app:proof_v2}

We now present a second possible approach for the proof of Eq.~\eqref{ineq:covCHSH}, which builds on the observation that the local bound $\frac{16}{7}$ on $\text{covCHSH}$ follows from just imposing the local bound $2$ on CHSH (which must be respected by any local distribution), and that probabilities are nonnegative.
More precisely, as previously we note that one can restrict to binary outputs $a,b = \pm 1$; as it turns out, it will then suffice to impose, together with $\text{CHSH} \le 2$, that $P(a,b|x,y) \ge 0$ for all $x,y,a,b$ satisfying $a b = (-1)^{xy+1}$.

\medskip

Instead of writing covCHSH as a function of the weights of deterministic distributions in a local decomposition, here we will view it directly as a function of the expectation values $\mean{A_xB_y}$, $\mean{A_x}$ and $\mean{B_y}$. Our goal is thus to maximize the value of covCHSH under the above constraints---which can also all be expressed in terms of those expectation values, by writing in particular $P(a,b|x,y) = \frac14(1+a \mean{A_x} + b \mean{B_y} + a b \mean{A_xB_y})$.

As previously, let us introduce KKT multipliers $\lambda$ and $\lambda_{xyab}$ (for the 8 combinations of $x,y,a,b$ such that $a b = (-1)^{xy+1}$), and define the Lagrangian
\ba \label{eq:lagrangian_v2}
&& \hspace{-4mm} L(\{\mean{A_xB_y}, \mean{A_x}, \mean{B_y}\}_{xy}, \lambda, \{\lambda_{xyab}\}_{xyab}) \nonumber \\[1mm]
&& = \text{covCHSH} + \lambda \, (2 - \text{CHSH}) \nonumber \\
&& \quad + \!\! \sum_{x,y,a,b: \,a b = (-1)^{xy+1}} \!\!\! \lambda_{xyab} \, P(a,b|x,y) . \hspace{-3mm}
\ea
The KKT necessary conditions for optimality are
\ba
&& \frac{\partial L}{\partial \mean{A_xB_y}} = (-1)^{xy} (1-\lambda) + {\textstyle \frac14} \! \sum_{a,b} \! \lambda_{xyab} \, a b = 0 \quad \ \forall x,y, \nonumber \\
&& \frac{\partial L}{\partial \mean{A_x}} = - \sum_y (-1)^{xy} \mean{B_y}+ {\textstyle \frac14} \! \sum_{y,a,b} \! \lambda_{xyab} \, a = 0 \qquad \forall x, \nonumber \\
&& \frac{\partial L}{\partial \mean{B_y}} = - \sum_x (-1)^{xy} \mean{A_x}+ {\textstyle \frac14} \! \sum_{x,a,b} \! \lambda_{xyab} \, b = 0 \qquad \forall y, \nonumber \\[1mm]
&& \lambda \, (2 - \text{CHSH}) = 0, \quad \lambda \ge 0, \quad 2 - \text{CHSH} \ge 0, \nonumber \\[1mm]
&& \lambda_{xyab} \, P(a,b|x,y) = 0, \quad \lambda_{xyab} \ge 0, \quad P(a,b|x,y) \ge 0, \nonumber \\
&& \hspace{33mm} \forall x,y,a,b:\, a b = (-1)^{xy+1}. \nonumber \\[-4mm] \label{eq:KKTcond_v2}
\ea
The first 3 lines define 8 linear equality constraints for the 8 multipliers $\lambda_{xyab}$, which can be solved and give
\ba
\lambda_{xyab} = 2 \big[ 1 - \lambda - a (-1)^y \mean{A_{\bar x}} - b (-1)^x \mean{B_{\bar y}} \big], \qquad \label{eq:lxyab}
\ea
with ${\bar x} = 1-x$ and ${\bar y} = 1-y$. We are then left with just the last couple of lines of Eq.~\eqref{eq:KKTcond_v2}, so that either $\lambda = 0$ or $\text{CHSH} = 2$, and for each $x,y,a,b$ such that $a b = (-1)^{xy+1}$, either $\lambda_{xyab} = 0$ or $P(a,b|x,y) = 0$. We can then consider the $2^{1+8} = 512$ corresponding cases separately, and solve for each case the resulting linear system of 9 equations for the 9 variables $\mean{A_xB_y}$, $\mean{A_x}$, $\mean{B_y}$ and $\lambda$.

Using Mathematica we found feasible solutions for those equations in $320$ of the $512$ cases. However, one still needs to check if these solutions satisfy the remaining inequality constraints in Eq.~\eqref{eq:KKTcond_v2}; only in $207$ of those cases could we find solutions that indeed satisfy all KKT conditions. Among all these solutions, we found that the maximal value of covCHSH was $\frac{16}{7}$, which again proves that this is the local bound of our covariance Bell inequality~\eqref{ineq:covCHSH}.
This value of $\frac{16}{7}$ was obtained for the $8$ solutions listed in Table~\ref{table:optim_expects}, which indeed define the same $8$ distributions as those listed (in the same order) in Table~\ref{table:optim_qs}.

\begin{table}
\centering
\begin{tabular}{cccccccccc}
& $\mean{A_0B_0}$ & $\mean{A_0B_1}$ & $\mean{A_1B_0}$ & $\mean{A_1B_1}$ & & $\mean{A_0}$ & $\mean{A_1}$ & $\mean{B_0}$ & $\mean{B_1}$ \\[1mm]
 \cline{2-10}
 \\[-4mm]
\ldelim\{{2}{3mm}[] & 1 & $\frac{3}{7}$ & $\frac{3}{7}$ & -$\frac{1}{7}$ && -$\frac{1}{7}$ & $\frac{3}{7}$ & -$\frac{1}{7}$ & $\frac{3}{7}$ \\[1mm]
 & 1 & $\frac{3}{7}$ & $\frac{3}{7}$ & -$\frac{1}{7}$ && $\frac{1}{7}$ & -$\frac{3}{7}$ & $\frac{1}{7}$ & -$\frac{3}{7}$ \\[1mm]
\ldelim\{{2}{3mm}[] & $\frac{1}{7}$ & $\frac{3}{7}$ & $\frac{3}{7}$ & -1 && $\frac{3}{7}$ & $\frac{1}{7}$ & -$\frac{3}{7}$ & -$\frac{1}{7}$ \\[1mm]
 & $\frac{1}{7}$ & $\frac{3}{7}$ & $\frac{3}{7}$ & -1 && -$\frac{3}{7}$ & -$\frac{1}{7}$ & $\frac{3}{7}$ & $\frac{1}{7}$ \\[1mm]
\ldelim\{{2}{3mm}[] & $\frac{3}{7}$ & $\frac{1}{7}$ & 1 & -$\frac{3}{7}$ && $\frac{3}{7}$ & -$\frac{1}{7}$ & -$\frac{1}{7}$ & -$\frac{3}{7}$ \\[1mm]
 & $\frac{3}{7}$ & $\frac{1}{7}$ & 1 & -$\frac{3}{7}$ && -$\frac{3}{7}$ & $\frac{1}{7}$ & $\frac{1}{7}$ & $\frac{3}{7}$ \\[1mm]
\ldelim\{{2}{3mm}[] & $\frac{3}{7}$ & 1 & $\frac{1}{7}$ & -$\frac{3}{7}$ && -$\frac{1}{7}$ & -$\frac{3}{7}$ & $\frac{3}{7}$ & -$\frac{1}{7}$ \\[1mm]
 & $\frac{3}{7}$ & 1 & $\frac{1}{7}$ & -$\frac{3}{7}$ && $\frac{1}{7}$ & $\frac{3}{7}$ & -$\frac{3}{7}$ & $\frac{1}{7}$
\end{tabular}
\caption{Optimal solutions of the KKT conditions of Eq.~\eqref{eq:KKTcond_v2} (together with $\lambda = \frac{5}{7}$ in all cases and $\lambda_{xyab}$ as given in Eq.~\eqref{eq:lxyab}), giving $\text{covCHSH} = \frac{16}{7}$ and defining the same distributions as in Table~\ref{table:optim_qs}.}
\label{table:optim_expects}
\end{table}

\subsection{Generalizing our proofs to other covariance Bell inequalities}

The two versions of the proof of inequality~\eqref{ineq:covCHSH} presented above could in principle be generalized to calculate the local bound of any covariance Bell inequality constructed from a linear Bell inequality with binary outcomes, following the recipe we suggest in the main text. However, as the number of inputs, and therefore the dimension of the relevant probability space, increase, the number of different cases to consider in the proof may become too large to be tractable in a reasonable time.

For the case of cov3322 with ternary inputs for instance, after reducing to binary outputs and taking into account normalization and non-signalling constraints, the relevant probability space is of dimension $\text{dim}_{\vec P} = 15$. Following the first approach for the proof above, local distributions can be decomposed onto the $2^6$ deterministic local distributions, and the number of different cases to consider (corresponding to the number of subsets of $d$ nonzero weights $q_k$, with $2 \le d \le \text{dim}_{\vec P}+1$) is $\sum_{d=2}^{16} \binom{2^6}{d} \ge 7 \times 10^{14}$, a number far too large for all cases to be considered separately.

One may hope that following our second approach for the proof, instead, may reduce the number of cases to be considered down to a tractable one. The first thing to check here would be which constraints (locality constraints in terms of CHSH or $I_{3322}$ inequalities, together with non-negativity of the probabilities constraints) are sufficient to impose in order to obtain the local bound on cov3322. We did not follow this approach any further, and leave as an open question, whether this would be tractable enough to provide an analytical proof of inequality~\eqref{ineq:cov3322}.

Instead, to obtain the local bound of inequality~\eqref{ineq:cov3322}, we resorted to numerical optimization, by optimizing over the $2^6 = 64$ weights $q_k$ in a local decomposition of $P = \sum_k q_k P_k^{\text{det}}$. Although the optimization problem is nonconvex and involves quite a few free parameters, we found that the result of the numerical optimization was stable enough when starting from different starting points, which makes us confident that the local bound of $\frac92$ in inequality~\eqref{ineq:cov3322} is indeed correct.

\medskip

We note, finally, that the same proof techniques as above do not work for Bell inequalities with Pearson correlators like inequality~\eqref{ineq:rCHSH}.
The first point to be noticed is that, as emphasized before, the local sets for binary outputs $a,b \in \{+1,-1\}$ and for more outputs $a,b \in [-1,+1]$ are not the same---indeed we found different local bounds for binary and ternary outputs. Even restricting to a fixed number of inputs and trying to follow our approach with KKT multipliers, the Lagrangians we would write in either of the two proof versions would not be nicely quadratic functions of the weights $q_k$ in a local decomposition of the form $P = \sum q_k P_k^{\text{det}}$ (as in Eq.~\eqref{eq:lagrangian_v1}), or of the expectation values $\mean{A_xB_y}$, $\mean{A_x}$ and $\mean{B_y}$ (as in Eq.~\eqref{eq:lagrangian_v2}), so that the KKT necessary conditions for linearity would not simplify to linear systems of equations. Instead of pursuing such an approach, we again resorted here to numerical optimizations. As before, the numerical results we obtained were stable enough to make us confident that the local bounds we give are correct.

\section{Tsirelson bounds and quantum violations for covCHSH and rCHSH}
\label{app:Tsirelson_bnd}

In this Appendix we give the proofs that the Tsirelson bounds for both covCHSH and rCHSH are the same as for CHSH, namely $2\sqrt{2}$. The proofs presented here are inspired by that of Landau (for rCHSH) in Ref.~\cite{landau}.

\subsection{Tsirelson bound for covCHSH}

First note that by Neumark's dilation theorem, any quantum correlation can be obtained by projective measurements on pure states in Hilbert spaces of large enough dimensions.

Consider a bipartite state $\ket{\psi} \in {\cal H}_{AB} = {\cal H}_A \otimes {\cal H}_B$ and projective measurement operators $\hat A_x$ for Alice and $\hat B_y$ for Bob (with the random variables $A_x$, $B_y$ corresponding to the results of those measurements). Let us then define the ket vectors $\ket{\alpha_x} = (\hat A_x \otimes \mathbbm{1}_B - \mean{\hat A_x}_\psi \mathbbm{1}_{AB}) \ket{\psi}$ and $\ket{\beta_y} = (\mathbbm{1}_A \otimes \hat B_y - \mean{\hat B_y}_\psi \mathbbm{1}_{AB}) \ket{\psi}$, where $\mathbbm{1}_X$ denotes the identity operator acting on the Hilbert space ${\cal H}_X$ and where $\mean{\hat A_x}_\psi = \bra{\psi} \hat A_x \otimes \mathbbm{1}_B \ket{\psi}$ and $\mean{\hat B_y}_\psi = \bra{\psi} \mathbbm{1}_A \otimes \hat B_y \ket{\psi}$. With these definitions, we have
\ba
&& \braket{\alpha_x}{\beta_y} \nonumber \\
&& = \bra{\psi} (\hat A_x \!\otimes\! \mathbbm{1}_B {-} \mean{\hat A_x}_\psi \mathbbm{1}_{AB}) (\mathbbm{1}_A \!\otimes\! \hat B_y {-} \mean{\hat B_y}_\psi \mathbbm{1}_{AB}) \ket{\psi} \nonumber \\
&& = \bra{\psi} \hat A_x \otimes \hat B_y \ket{\psi} - \mean{\hat A_x}_\psi \mean{\hat B_y}_\psi \nonumber \\
&& = \mean{A_xB_y} - \mean{A_x}\mean{B_y} = \text{cov}(A_x,B_y) \label{eq:proof_Ts_cov_AxBy}
\ea
and similarly,
\ba
\braket{\alpha_x}{\alpha_x} &=& \mean{A_x^2} - \mean{A_x}^2 = \sigma(A_x)^2, \nonumber \\
\braket{\beta_y}{\beta_y} &=& \mean{B_y^2} - \mean{B_y}^2 = \sigma(B_y)^2. \label{eq:proof_Ts_cov_sigmas}
\ea
Restricting to measurement operators with eigenvalues in $[-1,+1]$,\footnote{We note that the restriction to bounded operators with eigenvalues in $[-1,+1]$ is in fact not necessary in the proof, as it is sufficient to just impose $\sigma(A_x), \sigma(B_y) \le 1$ (for any state $\ket{\psi}$) for the Tsirelson bound on covCHSH to follow.} this implies in particular that
\ba
\big|\!\big| \bra{\alpha_x} \big|\!\big| &=& \sqrt{\braket{\alpha_x}{\alpha_x}} = \sigma(A_x) \le 1, \nonumber \\
\big|\!\big| \bra{\beta_y} \big|\!\big| &=& \sqrt{\braket{\beta_y}{\beta_y}} = \sigma(B_y) \le 1.
\ea

We thus obtain
\ba
&& \hspace{-7mm} \text{covCHSH} \nonumber \\
& = & \braket{\alpha_0}{\beta_0} + \braket{\alpha_0}{\beta_1} + \braket{\alpha_1}{\beta_0} - \braket{\alpha_1}{\beta_1} \nonumber \\
& \le & \big|\!\big| \bra{\alpha_0} \big|\!\big| \, \big|\!\big| \ket{\beta_0} + \ket{\beta_1} \big|\!\big| + \big|\!\big| \bra{\alpha_1} \big|\!\big| \, \big|\!\big| \ket{\beta_0} - \ket{\beta_1} \big|\!\big| \nonumber \\
& \le & \big|\!\big| \ket{\beta_0} + \ket{\beta_1} \big|\!\big| + \big|\!\big| \ket{\beta_0} - \ket{\beta_1} \big|\!\big| \nonumber \\
& \le & \sqrt{2 \big( \, \big|\!\big| \ket{\beta_0} + \ket{\beta_1} \big|\!\big|^2 + \big|\!\big| \ket{\beta_0} - \ket{\beta_1} \big|\!\big|^2 \, \big)} \nonumber \\
&& = 2 \sqrt{\big|\!\big| \ket{\beta_0} \big|\!\big|^2 + \big|\!\big| \ket{\beta_1} \big|\!\big|^2} \le 2 \sqrt{2}, \label{eq:proof_tsireslon_bnd_covCHSH}
\ea
which is necessarily satisfied by any quantum correlation. This bound can be reached by adequate measurements on a maximally entangled state, see Appendix~\ref{app:Tsirelson_bnd_explicit} below.

\medskip

Let us mention here that following further the proof of Ref.~\cite{landau}, one can actually also refine the characterization of the quantum set in terms of covariances. Indeed, defining the ket vectors $\ket{\gamma_0} = \ket{\alpha_0}, \ket{\gamma_1} = \ket{\alpha_1}, \ket{\gamma_2} = \ket{\beta_0}, \ket{\gamma_3} = \ket{\beta_1}$ and the matrix $\Gamma$ with coefficients $\Gamma_{ij} = \braket{\gamma_i}{\gamma_j}$ (for $0 \le i,j, \le 3$), one necessarily has that $\Gamma$ is positive semidefinite (as $\Gamma = M^\dagger M$ with $M = \sum_i \ketbra{\gamma_i}{i}$, where $\{\ket{i}\}$ is the computational basis). That is, taking Eqs.~\eqref{eq:proof_Ts_cov_AxBy}--\eqref{eq:proof_Ts_cov_sigmas} into account, one must have
\ba
&& \Gamma = \nonumber \\
&& \left(\!
\begin{array}{cccc}
\sigma(A_0)^2 & \Gamma_{01} & \text{cov}(\!A_0,\!B_0\!) & \text{cov}(\!A_0,\!B_1\!) \\
\Gamma_{01}^* & \sigma(A_1)^2 & \text{cov}(\!A_1,\!B_0\!) & \text{cov}(\!A_1,\!B_1\!) \\
\text{cov}(\!A_0,\!B_0\!) & \text{cov}(\!A_1,\!B_0\!) & \sigma(B_0)^2 & \Gamma_{23} \\
\text{cov}(\!A_0,\!B_1\!) & \text{cov}(\!A_1,\!B_1\!) & \Gamma_{23}^* & \sigma(B_1) ^2
\end{array}
\!\right) \ge 0. \nonumber \\[-3mm]
\ea
Given that the variances are upper-bounded by $1$, one also has $\tilde\Gamma \ge 0$, with the matrix $\tilde\Gamma$ obtained from $\Gamma$ above by replacing all diagonal terms by $1$. Now, it can be shown (following the same arguments as in Ref.~\cite{landau}, see also Ref.~\cite{NPA}) that the condition that there exists (real or complex) coefficients $\Gamma_{01}$ and $\Gamma_{23}$ such that $\tilde\Gamma \ge 0$ requires
\ba
&& \arcsin \text{cov}(A_0,B_0) + \arcsin \text{cov}(A_0,B_1) \nonumber \\[-1mm]
& & 
\ + \ \arcsin \text{cov}(A_1,B_0) - \arcsin \text{cov}(A_1,B_1) \ \leq \ \pi \qquad
\ea
(as well as the symmetrical inequalities obtained by permuting the measurement inputs and/or flipping the sign of the measurement outcomes), which must therefore necessarily be satisfied by quantum correlations, and strengthens the condition that $\text{covCHSH} \le 2 \sqrt{2}$.

\subsection{Tsirelson bound for rCHSH}
\label{app:Tsirelson_bnd_rCHSH}

To obtain the Tsirelson bound for rCHSH, we follow a very similar approach to the one above, considering now normalized ket vectors $\ket{\tilde\alpha_x} = \ket{\alpha_x} \! / |\!|\! \ket{\alpha_x} \!|\!|$ and $\ket{\tilde\beta_y} = \ket{\beta_y} \! / |\!|\! \ket{\beta_y} \!|\!|$ (if $\ket{\alpha_x}$ or $\ket{\beta_y}$ are null vectors, we define $\ket{\tilde\alpha_x}$ or $\ket{\tilde\beta_y}$ to be any unit vector orthogonal to all other ones).
With these, we now have
\ba
\braket{\tilde\alpha_x}{\tilde\beta_y} = \text{r}(A_x,B_y).
\ea
Because of the normalization of $\ket{\tilde\alpha_x}, \ket{\tilde\beta_y}$ and $\text{r}(A_x,B_y)$ (via the division by the variances $\sigma(A_x), \sigma(B_y)$), we do not need to restrict here to measurement operators with eigenvalues in $[-1,+1]$ (or such that $\sigma(A_x), \sigma(B_y) \le 1$).
With the same calculations as in Eq.~\eqref{eq:proof_tsireslon_bnd_covCHSH}, we obtain, for any choice of (bounded) measurement operators $\hat A_x, \hat B_y$,
\ba
\text{rCHSH} &=& \braket{\tilde\alpha_0}{\tilde\beta_0} + \braket{\tilde\alpha_0}{\tilde\beta_1} + \braket{\tilde\alpha_1}{\tilde\beta_0} - \braket{\tilde\alpha_1}{\tilde\beta_1} \nonumber \\
& & \le \ \cdots \ \le \ 2 \sqrt{2}.
\ea 
This bound can again be reached quantum mechanically, see below.

As just mentioned, here no restriction is imposed on the measurement outcomes (provided only that $\sigma(A_x), \sigma(B_y) < \infty$). The bound above also holds for local distributions (which can always be realized quantum mechanically) when measurement outcomes are not necessarily restricted to be binary; as we show explicitly in Appendix~\ref{app:Pearson_ternary}, it can also be reached locally with ternary outcomes. It is quite remarkable that the local and the Tsirelson bounds coincide in this case.

\medskip

Let us finally mention again that, as it was done in Ref.~\cite{landau} (see also Ref.~\cite{NPA}), a tighter characterization of the quantum set in terms of Pearson correlators can also be given: any quantum correlation satisfies
\ba
&& \arcsin \text{r}(A_0,B_0) + \arcsin \text{r}(A_0,B_1) \nonumber \\[-1mm]
& & 
\ + \ \arcsin \text{r}(A_1,B_0) - \arcsin \text{r}(A_1,B_1) \ \leq \ \pi
\ea
(as well as the symmetrical inequalities).

\subsection{Explicit quantum violations of inequalities~\eqref{ineq:covCHSH} and~\eqref{ineq:rCHSH}}
\label{app:Tsirelson_bnd_explicit}

The Tsirelson bounds for covCHSH and rCHSH can be reached by the same quantum mechanical correlations as those that reach the maximal quantum value of $\text{CHSH} = 2\sqrt{2}$. For instance, consider that Alice and Bob share a two-qubit maximally entangled state $\ket{\phi_+} = \frac{1}{\sqrt{2}}(\ket{0,0}+ \ket{1,1}) $ and perform the local measurements $\hat A_0 = \hat \sigma_{\textsc z}, \hat A_1 = \hat \sigma_{\textsc x}, \hat B_0 = \frac{\hat \sigma_{\textsc z}+\hat \sigma_{\textsc x}}{\sqrt{2}}$ and $\hat B_1 = \frac{\hat \sigma_{\textsc z}-\hat \sigma_{\textsc x}}{\sqrt{2}}$, where $\hat \sigma_{\textsc z}$ and $\hat \sigma_{\textsc x}$ denote the Pauli matrices. In that case the marginal expectation values $\mean{A_x}$ and $\mean{ B_y}$ vanish, so that covCHSH and rCHSH simply reduce to CHSH, and we indeed obtain covCHSH = rCHSH = CHSH = $2\sqrt{2}$.

By then rotating for instance Bob's measurement settings together around the $y$ axis of the Bloch sphere, one obtains the full circle drawn on Fig.~\ref{fig:geom}.

\section{Higher quantum violations of Ineq.~\eqref{ineq:covCHSH} with mixed states: a consequence of nonlinearity}
\label{app:quantum_mixed}

\begin{figure}
\begin{center}
\includegraphics[width=.95\linewidth]{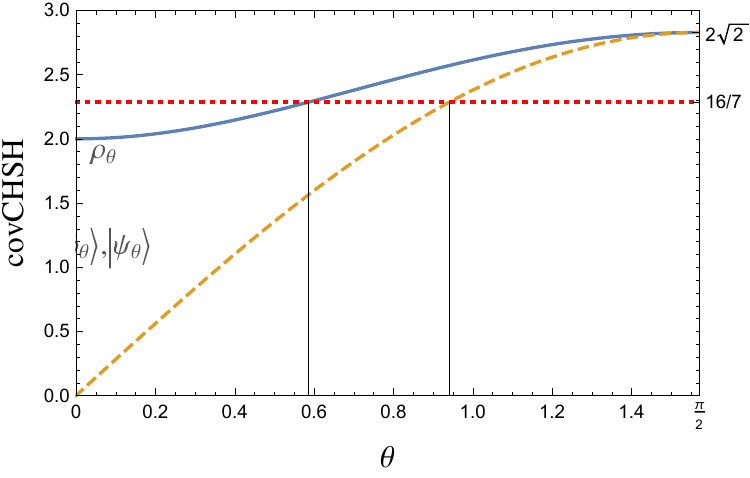}
\caption{Maximal value of the Bell expression covCHSH for the pure entangled states $\ket{\phi_{\theta}}$ and $\ket{\psi_{\theta}}$ (dashed orange curve), and for the mixture $\rho_\theta$ (solid blue curve). The horizontal dotted red line represents the local bound of the Bell inequality~\eqref{ineq:covCHSH}. In the interval $\theta \in \ ]\!\simeq 0.59 , \simeq 0.94]$, none of the two pure states violates the inequality, while the mixture does.}
	\label{fig:activation}
\end{center}
\end{figure}

In this Appendix we discuss an example illustrating the fact that the nonlinearity of the Bell inequality~\eqref{ineq:covCHSH} has nontrivial consequences also in the quantum case. Specifically, consider two pure entangled states of the form 
\ba
\ket{\phi_\theta} &=& \cos{\textstyle{\frac{\theta}{2}}} \ket{0,0} + \sin{\textstyle{\frac{\theta}{2}}} \ket{1,1}, \label{eq:phi_theta} \\
\ket{\psi_\theta} &=& \sin{\textstyle{\frac{\theta}{2}}} \ket{0,0} + \cos{\textstyle{\frac{\theta}{2}}} \ket{1,1}, \label{eq:psi_theta}
\ea
with $\theta \in \ ]0, \pi/2]$, and let us then define the equal mixture of these states:
\ba \label{eq:mixture}
\rho_{\theta} = \frac{1}{2} (\ket{\phi_\theta}\!\bra{\phi_\theta} + \ket{\phi_\theta}\!\bra{\phi_\theta}) \,.
\ea
We numerically estimated the largest value of the Bell expression~\eqref{eq:covCHSH}, optimizing over all local qubit measurements (including positive-operator valued measures) by Alice and Bob, for various values of the parameter $\theta$. The result is presented in Fig.~\ref{fig:activation}.
We found, up to numerical precision, that for the states $\ket{\phi_\theta}$ and $\ket{\psi_\theta}$ the value of covCHSH was maximal when using the Pauli measurements $\hat A_0 = \hat \sigma_{\textsc x}, \hat A_1 = \hat \sigma_{\textsc y}, \hat B_0 = \frac{\hat \sigma_{\textsc x}-\hat \sigma_{\textsc y}}{\sqrt{2}}$ and $\hat B_1 = \frac{\hat \sigma_{\textsc x}+\hat \sigma_{\textsc y}}{\sqrt{2}}$, giving $\text{covCHSH} = 2\sqrt{2} \sin \theta$ (orange dashed curve on Fig.~\ref{fig:activation}), while the optimal measurements on the state $\rho_\theta$ were found to be $\hat A_0 = \hat \sigma_{\textsc z}, \hat A_1 = \hat \sigma_{\textsc x}, \hat B_0 = \frac{\hat \sigma_{\textsc z}+\sin\theta\,\hat \sigma_{\textsc x}}{\sqrt{1+ \sin^2 \theta}}$ and $\hat B_1 = \frac{\hat \sigma_{\textsc z}-\sin\theta\,\hat \sigma_{\textsc x}}{\sqrt{1+ \sin^2 \theta}}$, giving $\text{covCHSH} = 2\sqrt{1+ \sin^2 \theta}$ (blue curve on Fig.~\ref{fig:activation}). This gives violations of inequality~\eqref{ineq:covCHSH} for $\theta > \arcsin \frac{16/7}{2\sqrt{2}} \simeq 0.94$ in the first case, and for $\theta > \arcsin \sqrt{(\frac{16/7}{2})^2-1} \simeq 0.59$ in the second case. We thus find that in the range of parameters $0.59 \lesssim \theta \lesssim 0.94$, none of the two pure states can individually violate the covariance Bell inequality~\eqref{ineq:covCHSH}, whereas the mixture~\eqref{eq:mixture} does violate it. This effect is possible only via the nonlinearity of the covariance Bell inequality. 

Note that the pure states $\ket{\phi_{\theta}}$ and $\ket{\psi_{\theta}}$ and the mixture $\rho$ give in fact the same correlation functions $\mean{A_xB_y}$ for any (projective) Pauli measurements $\hat A_x = \hat \sigma_{A_x}$ and $\hat B_y = \hat \sigma_{B_y}$ for Alice and Bob. The difference resides in the marginals $\mean{A_x}$ and $\mean{B_y}$, which are opposite for $\ket{\phi_{\theta}}$ and $\ket{\psi_{\theta}}$ and are therefore unbiased for the equal mixture $\rho_\theta$. 

Let us finally mention that we also investigated quantum violations of inequality~\eqref{ineq:covCHSH} using mixed entangled states of rank 3 and higher (up to dimension $6 \times 6$). We always found that the optimal violations were obtained with pure or rank-2 states. It thus seems that mixing more than 2 pure entangled states does not help increase the violations.

\section{Optimal local strategies and violations for our covariance Bell inequalities~\eqref{ineq:cov3322}--\eqref{ineq:rCHSH}}
\label{app:local_strats}

In this Appendix we present some local strategies that reach the local bounds of our Bell inequalities~\eqref{ineq:cov3322} and~\eqref{ineq:rCHSH}, and show how these can be violated quantum mechanically and by non-signalling correlations.
Recall that the local bounds were obtained numerically; however, we are confident that the numerics are trustworthy enough, and therefore that the local strategies we present below are indeed optimal.

\subsection{Covariance Bell inequality~\eqref{ineq:cov3322}}

\subsubsection{Optimal local strategies}


To obtain the local bound for cov3322 we ran a numerical optimization several times and conclude from the stability of the result that it must be $\frac{9}{2}$, as in inequality~\eqref{ineq:cov3322}. 

There are 4 different local distributions that give the value $\text{cov3322} = \frac{9}{2}$, and that decompose onto three deterministic correlations with weights $\frac38$, $\frac38$ and $\frac14$: namely, $P = \frac38 (+++/+++) + \frac38 (--+/--+) + \frac14 (-+-/-+-)$, $P = \frac38 (+++/++-) + \frac38 (--+/---) + \frac14 (+--/-++)$ (using here the notation $(A_0A_1A_2/B_0B_1B_2)$), and the other 2 ``opposite decompositions'', where all outputs are flipped.

\subsubsection{Violations of inequality~\eqref{ineq:cov3322}}

Inequality~\eqref{ineq:cov3322} can be violated by quantum correlations. For example, taking a two-qubit maximally entangled state $\ket{\phi_+} = \frac{1}{\sqrt{2}}(\ket{0,0}+ \ket{1,1}) $ and suitable measurements (see below), one can obtain cov3322$=5$, which is also the highest value for $I_{3322}$ (Eq.~\eqref{ineq:I3322}) obtainable with qubits and projective measurements~\cite{froissart,sliwa,CG} (note that in that case, all marginal terms vanish and cov3322 effectively reduces to $I_{3322}$).
We have performed an extensive numerical search, considering entangled states of dimension up to $5 \times 5$, and could not find any larger quantum violation than cov3322=5. It thus remains an open question, whether higher dimensional entangled states could lead to larger violations, as it is the case for the standard $I_{3322}$ inequality~\cite{tamas2}. 

Considering again the families of states $\ket{\phi_\theta}$, $\ket{\psi_\theta}$ and $\rho_\theta$ of Eqs.~\eqref{eq:phi_theta}--\eqref{eq:mixture}, the largest values of cov3322 we found (numerically, up to machine precision) are $\text{cov3322} = 5 \sin \theta$ for $\ket{\phi_\theta}$ or $\ket{\psi_\theta}$ (obtained e.g. for $\hat A_0 = \hat B_1 = \hat\sigma_{\textsc x}$, $\hat A_1 = \hat B_2 = \cos \frac{\pi}{3} \,\hat\sigma_{\textsc x} + \sin \frac{\pi}{3} \,\hat\sigma_{\textsc y}$, $\hat A_2 = -\hat B_0 = \cos \frac{2\pi}{3} \,\hat\sigma_{\textsc x} + \sin \frac{2\pi}{3} \,\hat\sigma_{\textsc y}$) and $\text{cov3322} = 4 + \sin^2 \theta$ for $\rho_\theta$ (obtained e.g. for $\hat A_0 = \hat B_0 = \sqrt{1-\frac{\sin^2 \theta}{4}} \,\hat\sigma_{\textsc z} + \frac{\sin \theta}{2} \,\hat\sigma_{\textsc x}$, $\hat A_1 = \hat B_1 = \sqrt{1-\frac{\sin^2 \theta}{4}} \,\hat\sigma_{\textsc z} - \frac{\sin \theta}{2} \,\hat\sigma_{\textsc x}$, $\hat A_2 = \hat B_2 = \hat\sigma_{\textsc x}$). These violate inequality~\eqref{ineq:cov3322} for $\theta > \arcsin \frac{9/2}{5} \simeq 1.12$ and $\theta > \arcsin \sqrt{\frac92-4} = \frac{\pi}{4}$, respectively. As it was the case with covCHSH, we thus find a range of parameters (for $\frac{\pi}{4} < \theta \lesssim 1.12$) for which neither $\ket{\phi_\theta}$ nor $\ket{\psi_\theta}$ violates inequality~\eqref{ineq:cov3322}, but their equal mixture does---which is a consequence of the nonlinearity of cov3322. Similarly to covCHSH this property does not seem to extend to mixtures of more than two pure states, that is, the violation of rank-3 states (or higher) does not appear to be larger than when mixing only two pure states. It is also interesting to note that within the families under consideration, cov3322 is violated by a smaller range of states than covCHSH. This is in contrast to the relationship between the standard CHSH and $I_{3322}$ linear Bell inequalities, where all entangled (possibly mixed) quantum states violating (a symmetry of) the former also violate the latter~\cite{CG}.%
\footnote{This can indeed be seen as follows: taking $\hat A_2 = \hat B_0 = -\openone$, the expression of cov3322 in Eq.~\eqref{ineq:I3322} simplifies to $I_{3322} = \mean{A_0B_1} + \mean{A_0B_2} + \mean{A_1B_1} - \mean{A_1B_2} +2 = \text{CHSH}_{01/12}+2$, where $\text{CHSH}_{01/12}$ is a symmetry of CHSH, Eq.~\eqref{eq:chsh}. Clearly if there exist measurement operators giving $\text{CHSH}_{01/12} > 2$, one then also gets a value $I_{3322} > 4$. Note that the same argument does not extend to covCHSH and cov3322: for the choice of $\hat A_2 = \hat B_0 = -\openone$, we just get $\text{cov3322} = \text{covCHSH}_{01/12} \le 4$, so that no violation of inequality~\eqref{ineq:cov3322} is possible with those measurement settings.}

\medskip

Note, finally, that the algebraic maximum value of $\text{cov3322} = 8$ can be attained by the non-signalling distributions such that $\mean{A_0 B_0} = \mean{A_0 B_1} = \mean{A_0 B_2} = \mean{A_1 B_0} = \mean{A_1 B_1} = - \mean{A_0 B_2} = \mean{A_2 B_0} = -\mean{A_2 B_1} = 1$ (with $\mean{A_2 B_2}$ remaining a free parameter) and with all marginals $\mean{A_x} = \mean{ B_y} = 0$. Such distributions also reach the maximal non-signalling value of $I_{3322} = 8$.

\subsection{Bell inequality~\eqref{ineq:rCHSH} for Pearson correlators}
\label{app:rCHSH}


\subsubsection{Optimal local strategies with binary outputs $a,b = \pm 1$}

For the case of binary outputs (taking e.g. $a,b = \pm 1$) we find, again numerically, that the local bound for rCHSH is $\frac{5}{2}$, as in inequality~\eqref{ineq:rCHSH}. This bound can be reached by the $8$ distributions that decompose as mixtures of the same triplets of deterministic distributions $P_{k_1}^{\text{det}}$, $P_{k_2}^{\text{det}}$, $P_{k_3}^{\text{det}}$ as those in Table~\ref{table:optim_qs} of Appendix~\ref{app:proofs} (i.e., those that allow one to reach the local bound of the covariance inequality~\eqref{ineq:covCHSH}), but taking here uniform mixtures, $P = \frac13 P_{k_1}^{\text{det}} + \frac13 P_{k_2}^{\text{det}} + \frac13 P_{k_3}^{\text{det}}$. E.g., for the first one: $P = \frac13 (++/++) + \frac13 (-+/--) + \frac13 (--/-+)$.

\subsubsection{Optimal local strategies with ternary outputs $a,b = +1,0,-1$}
\label{app:Pearson_ternary}

As emphasized before, in the case of Pearson correlators, considering binary outputs 
or a larger alphabet of possible outputs does make a difference.
As shown in Appendix~\ref{app:Tsirelson_bnd_rCHSH}, the value of rCHSH, for any choice of possible outputs, that can be reached locally is upper-bounded by $2\sqrt{2}$. As it turns out, this bound can be reached with ternary outputs $a,b = +1,0,-1$, for instance by the local distribution $P = \frac49 (++/+0) + \frac49 (+-/0+) + \frac19 (00/--)$.

Note that if any output variable $A_x$ or $B_y$ is deterministic, its variance is zero and by convention we took $\text{r}(A_x,B_y) = 0$. This implies that rCHSH defined in Eq.~\eqref{ineq:rCHSH} contains at most 2 nonzero Pearson correlators, which implies that $\text{rCHSH} \le 2$. Furthermore, if the two output variables of any party are both deterministic, then $\text{rCHSH} = 0$. 
Nevertheless, one can also locally reach a value of $\text{rCHSH}$ arbitrarily close to the bound $2 \sqrt{2}$ by certain distributions that get arbitrarily close to some deterministic ones: e.g., $P = (1-\epsilon) (++/+0) + \frac{\epsilon}{2}(+-/-+) + \frac{\epsilon}{2}(-+/--)$ with $0<\epsilon<1$ gives $\text{rCHSH} = 2(1+\sqrt{1-\epsilon})/\sqrt{2-\epsilon} \to 2\sqrt{2}$ as $\epsilon \to 0$ (note that Alice still has a binary output $\pm 1$ here; only Bob has a ternary output $+1,0,-1$, for one of his inputs only). Hence, in this case the value of $\text{rCHSH}$ cannot be used as a shared randomness witness.

\subsubsection{Violations of inequality~\eqref{ineq:rCHSH}}

As proven in Appendix~\ref{app:Tsirelson_bnd}, the Tsirelson bound for rCHSH is $2\sqrt{2}$, which can be reached by adequate (binary) projective measurements on a maximally 2-qubit entangled states. The quantum correlations thus obtained thus violate the inequality~\eqref{ineq:rCHSH} (with the local bound $\frac{5}{2}$, for the case where one restricts to binary outputs), but do not violate the analogous inequality for arbitrary outcomes, when the local bound also becomes $2\sqrt{2}$ (see above).

We note that for rCHSH, the nonlinearity of the Bell expression does not seem to exhibit the same interesting features as covCHSH and cov3322 presented in the previous appendices: the maximal value of a mixture of pure states appears, from our numerical investigations, to always be attained by one of the pure states. 

\medskip

Finally, note that algebraic maximum of rCHSH=4 can be obtained by a PR box (just as for CHSH and covCHSH).

\section{Minimal amount of shared randomness required to reach a given value of covCHSH}
\label{app:min_shared_randomness}

In this appendix we provide the details on how one can find the minimal amount of shared randomness---quantified here in terms of the Shannon entropy $H(\Lambda)$---required to locally reach a given value of covCHSH.

A general local model provides a decomposition for the probabilities $P(a,b|x,y)$ as in Eq.~\eqref{eq:local}. Our aim is thus to estimate the function
\ba
\min \!H_\Lambda^\text{gen}(c) \equiv \min_{\substack{ \text{local distributions~\eqref{eq:local}} \\ \text{giving } \text{covCHSH}=c }} H(\Lambda), \label{eqdef:minHgen}
\ea
for any value of $c$ between $0$ and $\frac{16}{7}$.

Rather than considering general local decompositions directly, it is in fact useful to first concentrate on the case of binary outputs $a,b = \pm 1$, and consider decompositions onto deterministic local response functions---that is, with $P_A (a|x,\lambda ), P_B (b|y,\lambda)$ equal to $0$ or $1$.
We shall thus first estimate the function
\ba
\min \!H_\Lambda^\text{det}(c) \equiv \min_{\substack{ \text{local distributions~\eqref{eq:local}} \\ \text{with } a,b = \pm 1 \text{ and } P_A, P_B = 0, 1 \\ \text{giving } \text{covCHSH}=c }} H(\Lambda), \quad \label{eqdef:minHdet}
\ea
which clearly satisfies $\min \!H_\Lambda^\text{det}(c) \ge \min \!H_\Lambda^\text{gen}(c)$. We will then prove that these two functions in fact coincide: to reach the minimal amount of shared randomness for a given value of covCHSH, it is indeed enough to just consider binary outputs $\pm 1$, and decompositions onto deterministic local response functions.

\subsection{Local decompositions onto deterministic response functions with binary outputs $\pm 1$}

Let us first note that for binary inputs and binary outputs, there is only a finite number ($16$) of local deterministic strategies for Alice and Bob together. Even though the local decomposition~\eqref{eq:local} with deterministic response functions may involve more than $16$ different values for $\lambda$ (and even continuous values, in an appropriate limit), clearly the entropy $H(\Lambda)$ is minimized when the different values of $\lambda$ that define the same deterministic strategies are grouped together. This implies that it is sufficient to consider here local decompositions of the form $P = \sum_k q_k P_k^{\text{det}}$ onto the $16$ different deterministic distributions $P_k^{\text{det}}$. The optimization problem in Eq.~\eqref{eqdef:minHdet} then reduces to optimize the $16$ weights $q_k$ under the constraints $q_k \ge 0$, $\sum_k q_k = 1$ and $\text{covCHSH}(\{q_k\}_k) = c$.

We performed this optimization numerically, for various values of $c = \text{covCHSH}$. Our results are shown on Fig.~\ref{fig:entropy} (the function $\min \!H_\Lambda^\text{det}$ coincides with $\min H(\Lambda)$ shown there). We repeated the numerical optimization several times, starting with different random starting points; the consistency of the results we obtained make us confident that we indeed reached the global minima in Eq.~\eqref{eqdef:minHdet}.

What we found (up to numerical precision) is that for $c = \text{covCHSH} \le 2$, the minimum in~\eqref{eqdef:minHdet} is reached by mixing only $2$ deterministic local distributions with nonzero weights: for instance, taking $P = q (++/++) + (1{-}q) (--/--)$ gives $\text{covCHSH} = 8q(1{-}q)$, $H(\Lambda) = H(\{q,1{-}q\}) = -q \log_2 q - (1{-}q) \log_2(1{-}q)$, and leads (for $0 \le c \le 2$) to $\min \!H_\Lambda^\text{det}(c) = h_2(\sqrt{1-c/2})$, with the binary entropy function $h_2(x) = -\frac{1+x}{2}\log_2\frac{1+x}{2}-\frac{1-x}{2}\log_2\frac{1-x}{2}$.

For $2 < c = \text{covCHSH} \le \frac{16}{7} \simeq 2.29$, we found that the minimum  in~\eqref{eqdef:minHdet} is reached by mixing now $3$ deterministic local distributions with nonzero weights: for instance, with distributions of the form $P = q_1 (++/++) + q_2 (-+/--)  + q_3 (--/-+)$. That gives $\text{covCHSH} = (1-q_1)(1+7q_1)-(q_2-q_3)^2$, $H(\Lambda) = H(\{q_1,q_2,q_3\})$. For a given value of $\text{covCHSH} = c$, that reduces to optimizing over just one parameter, say $q_1$, as $q_2$ and $q_3$ are then given by $q_{2,3} = \frac{1-q_1\pm\sqrt{(1-q_1)(1+7q_1)-c}}{2}$ (which requires $\frac{3-\sqrt{16-7c}}{7} \le q_1 \le \frac{3+\sqrt{16-7c}}{7}$, so that all $q_k$ are between $0$ and $1$). For $c \gtrsim 2.27$ we find that the optimal is obtained for $q_1 = \frac{3+\sqrt{16-7c}}{7}$ (and $q_2=q_3=\frac{1-q_1}{2}$); for $c \lesssim 2.27$ we cannot give an analytical solution (however, a good approximation seems to be $q_1 \simeq \frac12 - 0.01 (c - 2)$).

\medskip

We note that as can clearly be seen on Fig.~\ref{fig:entropy}, $\min \!H_\Lambda^\text{det}$ is a (strictly) increasing function of $c = \text{covCHSH}$.

\subsection{General local decompositions}

For a general local decomposition of the form~\eqref{eq:local}, one gets
\ba
&& \hspace{-1mm} \text{covCHSH} \nonumber \\
&& = \! \int \!\! \rho(\lambda) \big[ \mean{A_0}_\lambda \; (\mean{B_0}_\lambda {-} \mean{B_0} {+} \mean{B_1}_\lambda {-} \mean{B_1}) \nonumber \\[-2mm]
&& \qquad \qquad + \mean{A_1}_\lambda \; (\mean{B_0}_\lambda {-} \mean{B_0} {-} \mean{B_1}_\lambda {+} \mean{B_1}) \big] d\lambda \qquad
\ea
with $\mean{A_x}_\lambda = \sum_{a} a \, P(a|x,\lambda)$ and $\mean{B_y}_\lambda = \sum_{b} b \, P(b|y,\lambda)$ (so that $\mean{A_x} = \int \rho(\lambda) \mean{A_x}_\lambda d\lambda$ and similarly for $\mean{B_y}$).

Looking at this expression, it is clear that for a given distribution $\rho(\lambda)$ (and therefore a given value of $H(\Lambda)$) and some given response functions of Bob, the maximal value of covCHSH is obtained when $|\mean{A_0}_\lambda| = 1$, i.e., when Alice uses deterministic response functions with output $\pm 1$ (specifically, Alice's optimal response function is $A_x(\lambda) = \text{sign}[\mean{B_0}_\lambda {-} \mean{B_0} {+} (-1)^x (\mean{B_1}_\lambda {-} \mean{B_1})]$).

A similar reasoning can be followed for Bob. We then conclude that for a given value of $H(\Lambda)$, the maximal value of covCHSH is obtained when Alice and Bob use deterministic response functions with outputs $a,b = \pm 1$: essentially, local randomness (or considering other possible outputs in $[-1,+1]$) does not help increase the value of covCHSH.

\medskip

Note, however, that the optimization problem in Eq.~\eqref{eqdef:minHgen} is subtly different: rather than maximizing covCHSH for a given value of $H(\Lambda)$, we want to mimimize $H(\Lambda)$ for a given value of covCHSH. To conclude here we will use the previous observation, that $\min \!H_\Lambda^\text{det}(c)$ is an increasing function of $c=\text{covCHSH}$.

Consider indeed a (general) local decomposition of a distribution $P$ with a given amount of shared randomness $H(\Lambda) = h$, and which gives some value $\text{covCHSH} = c$. From the above reasoning, it follows that there exists another distribution $P'$ using the same shared random variable $\Lambda$ (hence, with the same value $H(\Lambda) = h$) but now decomposed onto deterministic local response functions with binary outputs $\pm 1$, that gives a value $\text{covCHSH} = c' \ge c$. This then implies that $\min \!H_\Lambda^\text{det}(c') \le h$ and, because $\min \!H_\Lambda^\text{det}$ is an increasing function, $\min \!H_\Lambda^\text{det}(c) \le h$ as well.
Hence, any general local decomposition that gives some value $\text{covCHSH} = c$ necessarily  satisfies $H(\Lambda) \ge \min \!H_\Lambda^\text{det}(c)$. This implies that $\min \!H_\Lambda^\text{gen}(c) \ge \min \!H_\Lambda^\text{det}(c)$---and therefore, $\min \!H_\Lambda^\text{gen}(c) = \min \!H_\Lambda^\text{det}(c)$---with $\min \!H_\Lambda^\text{det}(c)$ estimated above.

\end{document}